\documentclass[twocolumn,showpacs,preprintnumbers,amsmath,amssymb]{revtex4}

\usepackage{psfig}
\usepackage{stmaryrd} 
\usepackage{graphicx} 

\begin{document}


\title{Information Theory --- The Bridge Connecting Bounded Rational Game
Theory and Statistical Physics}

\author{David H. Wolpert}
\affiliation{NASA Ames Research Center, Moffett Field, CA, 94035, USA\\
{\tt dhw@email.arc.nasa.gov}}

\begin{abstract}
A long-running difficulty with conventional game theory has been how
to modify it to accommodate the bounded rationality of all real-world
players. A recurring issue in statistical physics is how best to
approximate joint probability distributions with decoupled (and
therefore far more tractable) distributions. This paper shows that the
same information theoretic mathematical structure, known as Product
Distribution (PD) theory, addresses both issues. In this, PD theory
not only provides a principled formulation of bounded rationality and
a set of new types of mean field theory in statistical physics. It
also shows that those topics are fundamentally one and the same.
\end{abstract}

\pacs{89.20.-a, 89.75.-k , 89.75.Fb}


\maketitle              

\nocite{chjo02, kono03_short}

\section{Introduction} \label{sec:intro} 

In noncooperative game theory, one has a set of $N$ players, each
choosing its strategy $x_i$ independently, by sampling a distribution
$q_i(x_i)$ over those strategies. Each player $i$ also has her own
utility function $g_i(x)$, specifying how much reward she gets for
every possible joint-strategy $x$ of all $N$ players. Let
$q_{(i)}(x_{(i)})$ mean the joint probability distribution of all
players other than $i$, i.e., $\prod_{j \ne i} q_j(x_j)$.  Then the
``goal'' of each player $i$ is to set $q_i$ to so that, conditioned on
$q_{(i)}$, the expected value of $i$'s utility is as high as possible.

Conventional game theory assumes each player $i$ is ``fully
rational'', able to solve for that optimal $q_i$, and that she then
uses that distribution. It is primarily concerned with analyzing the
such equilibria of the game \cite{futi91,baol99,osru94,auha92}. In the
real world, this assumption of full rationality almost never holds,
whether the players are humans, animals, or computational agents
\cite{axel84,sale97,neym85,fule98,arth94,bosh97,nasm01,tvka92,kahn03}. This
is due to the cost of computation of that optimal distribution, if
nothing else. This real-world {\bf{bounded rationality}} is one of the
major impediments to applying conventional game theory in the real
world.

This paper shows how Shannon's information theory
\cite{coth91,mack03,jabr03} provides a principled way to modify
conventional game theory to accommodate bounded rationality. This is
done by following information theory's prescription that, given only
partial knowledge concerning the distributions the players are using,
we should use the Maximum Entropy (Maxent) principle to infer those
distributions. Doing so results in the principle that the bounded
rational equilibrium is the minimizer of a certain set of coupled
Lagrangian functions of the joint distribution, $q(x) = \prod_i
q_i(x_i)$. This mathematical structure is a special instance of
Product Distribution (PD) theory
\cite{wolp03,wolp04b,mabi04,lewo04,biwo04,arth94}.

In addition to showing how to formulate bounded rationality, 
PD theory provides many other advantages to game theory. Its
formulation of bounded rationality explicitly includes a term that, in
light of information theory, is naturally interpreted as a cost of
computation. PD theory also seamlessly accommodates multiple utility
functions per player. It also provides many powerful techniques for
finding (bounded rational) equilibria, and helps address the issue of
multiple equilibria. Another advantage is that by changing the
coordinates of the underlying space of joint moves $x$, the same
mathematics describes a type of bounded rational cooperative game
theory, in which the moves of the players are transformed into
contracts they all offer one another.

Perhaps the most succinct and principled way of deriving statistical
physics is as the application of the Maxent principle. In this
formulation, the problem of statistical physics is cast as how best to
infer the probability distribution over a system's states when one's
prior knowledge consists purely of the expectation values of certain
functions of the system's state \cite{jayn57,jabr03}. For example,
this prescription says we should infer that the probability
distribution $p$ governing the system is the Boltzmann distribution
when our prior knowledge is the system's expected energy. This is
known as the ``canonical ensemble''. Other ensembles arise when other
expectation values are added to one's prior knowledge. In particular,
if the number of particles in the system is uncertain, but one knows
its expectation value, one arrives at the ``grand canonical
ensemble''.

One major difficulty with working with these ensembles is that under
them the particles of the system are statistically coupled with one
another. For high-dimensional systems, this can make statistical
physics calculations very difficult. Accordingly, a large body of work
has been produced under the rubric of Mean Field (MF) theory, in which
the ensemble is approximated with a distribution in which the
particles are independent \cite{opsa01}. In an MF approximation, a
product distribution $q$ governs the joint state of the particles ---
just as a product distribution governs the joint strategy of the
players in a game.

MF approximations are usually derived in an ad hoc manner. The
principled way to derive a MF approximation (or any other kind) to a
particular ensemble is to specify a distance measure saying how close
two probability distributions are, and then solve for the $q$ that is
closest to the distribution being approximated, $p$. To do this one
needs to specify the distance measure. How best to measure distances
between probability distributions is a topic of ongoing controversy and
research \cite{woma04a}. The most common way to do so is with the
infinite limit log likelihood of data being generated by one
distribution but misattributed to have come from the other. This is
known as the Kullback-Leibler (KL) distance
\cite{coth91,duha00,mack03}. It is far from being a metric. In
particular, it is not symmetric under interchange of the two
distributions being compared.

It turns out that the simplest MF theories minimize the KL distance
from $q$ to $p$.  However it can be argued it is the KL distance from
$p$ to $q$ that is the most appropriate measure, not the KL distance
from $q$ to $p$. Using that distance, the optimal $q$ is a new kind of
approximation not usually considered in statistical physics.

For the canonical ensemble, the type of KL distance arising in simple
MF theories turns out to be identical to the maxent Lagrangian arising
in bounded rational game theory. This shows how bounded rational
(independent) players are formally identical to the particles in the
MF approximation to the canonical ensemble. Under this identification,
the moves of the players play the roles of the states of the
particles, and particle energies are translated into player utilities.
The coordinate transformations which in game theory result in
cooperative games are, in statistical physics, techniques for more
allowing the canonical ensemble to be more accurately approximated
with a product distribution.

This identification raises the potential of transferring some of the
powerful mathematical techniques that have been developed in the
statistical physics community to the analysis of noncooperative game
theory. In also suggests translating some of the other ensembles of
statistical physics to game theory, in addition to the canonical
ensemble. As an example, in the grand canonical ensemble the number of
particles is variable, which, after a MF approximation, corresponds to
having a variable number of players in game theory. Among other
applications, this provides us with a new framework for analyzing
games in evolutionary scenarios, different from evolutionary game
theory.

In the next section noncooperative game theory and information theory
are cursorily reviewed. Then bounded rational game theory is derived,
and its many advantages are discussed. The following section starts
with a cursory review of the information-theoretic derivation of
statistical physics. After that is a discussion of the two kinds of KL
distance and the MF theories they induce, and a discussion of
coordinate systems. This section also includes a discussion on
translating a MF version of the grand canonical ensemble into a new
kind of evolutionary game theory.

As discussed in the physics section, the maxent Lagrangian and
associated Boltzmann solution at the core of this paper has been
investigated for an extremely long time in the context of
many-particle systems.  Considered in the context of a 2-player game
with nature, the Boltzmann solution has also been studied for many
years in the reinforcement learning community
~\cite{kali96,suba98}. Related work has considered it in the context
of ``mechanism design'' of many players, i.e., in the context of
designing the utility functions of the players to induce them to
maximize social welfare ~\cite{wotu99a,wotu01a,wotu02a,wolp03a}.

It turns out that independent of the work reported in this paper, the
maxent Lagrangian and its Boltzmann solution has been been muted in
the context of game theory ~\cite{shar04,fukr93,fule98}. However its
motivation in that work is somewhat {\it {ad hoc}}. In that work there
is no use of information theory nor discussion of the relation between
bounded rational game theory and mean field theory. There is also no
relation of the maxent Lagrangian to the cost of computation, multiple
cost functions, rationality operators, or the kinds of alternatives to
evolutionary game theory discussed below. Nor is there discussion of
semi-coordinate transformations and their relation to cooperative game
theory.

Finally, it's important to note that PD theory also has many
applications in science beyond those considered in this paper. For
example, see ~\cite{mabi04,lewo04,aiwo04,wobi04,anbi04} for work
relating the maxent Lagrangian to distributed control and to
distributed optimization. See ~\cite{wobi04} for algorithms for
speeding up convergence to bounded rational equilibria. Some of those
algorithms are related to simulated and deterministic annealing
~\cite{duha00}.  In ~\cite{wolp04b} others of those algorithms are
related to Stackelberg games, and more generally to the problem of
finding the optimal control hierarchy for team of players with a
common goal, i.e., finding an optimal organization chart. See also
~\cite{wole04,wolp04c,wolp04d} for work showing, respectively, how to
use PD theory to improve Metropolis-Hastings sampling, how to relate
it to the mechanism design work in
~\cite{wotu99a,wotu01a,wotu02a,wolp03a}, and how to extend it to
continuous move spaces and time-extended strategies.

\section{PD theory as Bounded Rational Noncooperative Game Theory}

This section motivates PD theory as a way of addressing several of the
shortcomings of conventional noncooperative game theory. 

\subsection{Review of noncooperative game theory}

In noncooperative game theory one has a set of $N$
{\bf{players}}. Each player $i$ has its own set of allowed {\bf{pure
strategies}}. A {\bf {mixed strategy}} is a distribution $q_i(x_i)$
over player $i$'s possible pure strategies.  Each player $i$ also has
a {\bf{utility function}} $g_i$ that maps the pure strategies adopted
by all $N$ of the players into the real numbers. So given mixed
strategies of all the players, the expected utility of player $i$ is
$E(g_i) = \int dx \; \prod_j q_j(x_j) g_i(x)$ \footnote{Throughout
this paper, the integral symbol will be interpreted in the appropriate
measure-theoretic terms, e.g., as Lebesgue integrals, point-sums,
etc.}.

This basic framework can be elaborated to model many interactions
between biological organisms, and in particular between human
beings. These interactions range from simple abstractions like the
famous prisoner's dilemma to iterated games like chess, to
international relations ~\cite{grei99,futi91,baol99}.

Much of noncooperative game theory is concerned with {\bf{equilibrium
concepts}} specifying what joint-strategy one should expect to result
from a particular game. In particular, in a {\bf{Nash equilibrium}} every
player adopts the mixed strategy that maximizes its expected utility,
given the mixed strategies of the other players. More formally,
$\forall i, q_i = {\mbox{argmax}}_{q'_i} \int dx \; q'_i \prod_{j \ne i}
q_j(x_j) \; g_i(x)$. 

Several very rich fields have benefited from a close relationship with
noncooperative game theory. Particular examples are evolutionary game
theory (in which the set of $N$ players is replaced by an infinite set
of reproducing organisms) and cooperative game theory (in which
players choose which {\bf{coalitions}} of other players to join)
~\cite{mead98,auha92}. Game theory as a whole is also closely related
to economics, in particular the field of mechanism design, which is
concerned with how to induce the set of players to do adopt a socially
desirable joint-strategy ~\cite{zlro99,futi91,niro01,wotu02c}.

\subsection{Problems with conventional noncooperative game theory}

A number of objections to the Nash equilibrium concept have been
resolved. In particular, it was Nash who proved that every game has at
least one Nash equilibrium if one expands the realm of discourse to
include mixed strategies. (The same is not true for pure strategies.)
Other objections have been more or less resolved through numerous
{\bf{refinements}} of the Nash equilibrium concept.

However there are several major problems with the concept that are
still outstanding. One of them is the possible multiplicity of
equilibria; this multiplicity means the Nash equilibrium concept
cannot be used to specify the joint strategy that is actually adopted
in a real world game. (Some refinements of the Nash equilibrium
concept attempt to address this problem, though none has succeeded.)
Another problem is that while calculating Nash equilibria is
straightforward in many simple games (e.g., 2 players in a zero-sum
game), calculating them in the general case can be a very difficult
computational multi-criteria optimization problem. Yet another problem
is that there is no general way to extend the concept to allow each
player to have multiple utility functions.

However perhaps the major problem with the Nash equilibrium concept is
its assumption of {{\bf full rationality}}. This is the assumption that
every player $i$ can both calculate what the strategies $q_{j \ne i}$
will be and then calculate its associated optimal distribution. In
other words, it is the assumption that every player will calculate the
entire joint distribution $q(x) = \prod_j q_j(x_j)$. If for no other
reasons than computational limitations of real humans, this assumption
is essentially untenable. This problem is just as severe if one allows
statistical coupling among the players ~\cite{auma87,futi91}.

A large body of empirical lore has been generated characterizing the
bounded rationality of humans. Similarly much has been
learned about the empirical behavior of (bounded rational) machine
learning computer algorithms playing games with one another
~\cite{axel84,nasm01}. None of this work has resulted in a full mathematical
theory of bounded rationality however.

There have also been numerous theoretical attempts to incorporate
bounded rationality into noncooperative game theory by modifying the
Nash equilibrium concept. Some of them assume essentially that every
player's mixed strategy is its Nash-optimal strategy with some form of
noise superimposed ~\cite{auha92}. Others explicitly model the humans,
typically as computationally limited automata, and assume the automata
perform optimally subject to those computational limitations
~\cite{fule98}. Both approaches, while providing insight, are very
{\it{ad hoc}} as models of games involving real-world organisms or
real-world (i.e., non-trivial) machine learning algorithms.

The difficulty of calculating equilibria is addressed in the sections
below on solving for the distributions of PD theory.  The rest of this
section shows how information theory can be used to extend game theory
to avoid its other shortcomings. Finally, the sections after this one
present some other extensions of game theory, in particular to allow
for a variable number of players. (Games with variable number of
players arise in many biological scenarios as well as economic ones.)

\subsection{Review of the maximum entropy principle}

Shannon was the first person to realize that based on any of several
separate sets of very simple desiderata, there is a unique real-valued
quantification of the amount of syntactic information in a
distribution $P(y)$. He showed that this amount of information is (the
negative of) the Shannon entropy of that distribution, $S(P) = -\int
dy \; P(y) ln[\frac{P(y)}{\mu(y)}]$ \footnote{$\mu$ is an {\it{a
priori}} measure over $y$, often interpreted as a prior probability
distribution. Unless explicitly stated otherwise, in this paper we
will always assume it is uniform, and not write it explicitly. See
\cite{jayn57,jabr03,coth91}.}.

So for example, the distribution with minimal information is the one
that doesn't distinguish at all between the various $y$, i.e., the
uniform distribution. Conversely, the most informative distribution is
the one that specifies a single possible $y$. Note that for a product
distribution, entropy is additive, i.e., $S(\prod_i q_i(y_i)) = \sum_i
S(q_i)$.

Say we given some incomplete prior knowledge about a distribution
$P(y)$. How should one estimate $P(y)$ based on that prior knowledge?
Shannon's result tells us how to do that in the most conservative way:
have your estimate of $P(y)$ contain the minimal amount of extra
information beyond that already contained in the prior knowledge about
$P(y)$.  Intuitively, this can be viewed as a version of Occam's
razor. This approach is called the maximum entropy (maxent)
principle. It has proven extremely useful in domains ranging from
signal processing to image processing to supervised learning
~\cite{mack03}.

\subsection{Maxent Lagrangians}

Much of the work on equilibrium concepts in game theory adopts the
perspective of an external observer of a game. We are told something
concerning the game, e.g., its utility functions, information sets,
etc., and from that wish to predict what joint strategy will be
followed by real-world players of the game.  Say that in addition to
such information, we are told the expected utilities of the players.
What is our best estimate of the distribution $q$ that generated those
expected utility values? By the maxent principle, it is the
distribution with maximal entropy, subject to those expectation
values.

To formalize this, for simplicity assume a finite number of players
and of possible strategies for each player. To agree with the
convention in other fields, from now on we implicitly flip the sign of
each $g_i$ so that the associated player $i$ wants to minimize that
function rather than maximize it. Intuitively, this flipped $g_i(x)$
is the ``cost" to player $i$ when the joint-strategy is $x$, rather
than its utility then.

Then for prior knowledge that the expected utilities of the players
are given by the set of values \{$\epsilon_i$\}, the maxent estimate
of the associated $q$ is given by the minimizer of the Lagrangian
\begin{eqnarray}
L(q) &\equiv& \sum_i \beta_i [E_q(g_i) - \epsilon_i]
- S(q) \notag \\
&=&  \sum_i \beta_i [\int dx \; \prod_j q_j(x_j)
g_i(x) -  \epsilon_i] - S(q)
\label{eq:firstlag}
\end{eqnarray}
\noindent where the subscript on the expectation value indicates that
it evaluated under distribution $q$, and the \{$\beta_i$\} are
Lagrange parameters implicitly set by the constraints on the expected
utilities \footnote{Throughout this paper the terms in any Lagrangian
that restrict distributions to the unit simplices are implicit. The
other constraint needed for a Euclidean vector to be a valid
probability distribution is that none of its components are
negative. This will not need to be explicitly enforced in the
Lagrangian here.}. 

Solving, we find that the mixed strategies minimizing the Lagrangian
are related to each other via
\begin{equation}
q_i(x_i) \propto e^{-E_{q_{(i)}}(G \mid x_i)} 
\label{eq:boltzG}
\end{equation}
\noindent where the overall proportionality constant for each $i$ is
set by normalization, and $G \equiv \sum_i \beta_i g_i$, and the
subscript $q_{(i)}$ on the expectation value indicates that it is
evaluated according the distribution $\prod_{j \ne i} q_j$.  In
Eq.~\ref{eq:boltzG} the probability of player $i$ choosing pure
strategy $x_i$ depends on the effect of that choice on the utilities
of the other players. This reflects the fact that our prior knowledge
concerns all the players equally.

If we wish to focus only on the behavior of player $i$, it is
appropriate to modify our prior knowledge.  To see how to do this,
first consider the case of maximal prior knowledge, in which we know
the actual joint-strategy of the players, and therefore all of their
expected costs. For this case, trivially, the maxent principle says we
should ``estimate" $q$ as that joint-strategy (it being the $q$ with
maximal entropy that is consistent with our prior knowledge). The same
conclusion holds if our prior knowledge also includes the expected
cost of player $i$.

Now modify this maximal set of prior knowledge by removing from it
specification of player $i$'s strategy. So our prior knowledge is
the mixed strategies of all players other than $i$, together with
player $i$'s expected cost. We can incorporate the prior knowledge of
the other players' mixed strategies directly into our Lagrangian,
without introducing Lagrange parameters. That {\bf{maxent
Lagrangian}} is
\begin{eqnarray*}
L_i(q_i) &\equiv& \beta_i [\epsilon_i - E(g_i)] - S_i(q_i) \nonumber \\
&=& \beta_i [\epsilon_i - \int dx \; \prod_j q_j(x_j) g_i(x)]
- S_i(q_i) 
\label{eq:maxentlag}
\end{eqnarray*}
\noindent with solution given by a set of coupled {\bf{Boltzmann
distributions}}:
\begin{equation}
q_i(x_i)  \propto e^{-\beta_i E_{q_{(i)}}(g_i \mid x_i)} .
\label{eq:boltzpr}
\end{equation}
\noindent Following Nash, we can use Brouwer's fixed point theorem to
establish that for any non-negative values \{$\beta$\}, there must
exist at least one product distribution given by the product of these
Boltzmann distributions (one term in the product for each $i$).

The first term in $L_i$ is minimized by a perfectly rational
player. The second term is minimized by a perfectly {\it{irrational}}
player, i.e., by a perfectly uniform mixed strategy $q_i$. So
$\beta_i$ in the maxent Lagrangian explicitly specifies the balance
between the rational and irrational behavior of the player. In
particular, for $\beta \rightarrow \infty$, by minimizing the
Lagrangians we recover the Nash equilibria of the game. More formally,
in that limit the set of $q$ that simultaneously minimize the
Lagrangians is the same as the set of delta functions about the Nash
equilibria of the game. The same is true for Eq.~\ref{eq:boltzG}.

Eq.~\ref{eq:boltzG} is just a special case of Eq.~\ref{eq:boltzpr},
where all player's share the same cost function $G$. (Such games are
known as {\bf{team games}}.) This relationship reflects the fact that
for this case, the difference between the maxent Lagrangian and the
one in Eq.~\ref{eq:firstlag} is independent of $q_i$. Due to this
relationship, our guarantee of the existence of a solution to the set
of maxent Lagrangians implies the existence of a solution of the form
Eq.~\ref{eq:boltzG}.

Typically players aren't close to perfectly self-defeating. Almost
always they will be closer to minimizing their expected cost than
maximizing it. For prior knowledge consistent with such a case, the
$\beta_i$ are all non-negative. 

Finally, our prior knowledge often will not consist of exact
specification of the expected costs of the players, even if that
knowledge arises from watching the players make their moves. Such
other kinds of prior knowledge are addressed in several of the
following subsections.

\subsection{Alternative interpretations of Lagrangians}

There are numerous alternative interpretations of these results. For
example, change our prior knowledge to be the entropy of each player
$i$'s strategy, i.e., how unsure it is of what move to make.  Now we
cannot use information theory to make our estimate of $q$.  Given that
players try to minimize expected cost, a reasonable alternative is
to predict that each player $i$'s expected cost will be as small as
possible, subject to that provided value of the entropy and the other
players' strategies. The associated Lagrangians are $\alpha_i[S(q_i) -
\sigma_i] - E(g_i)$, where $\sigma_i$ is the provided entropy
value. This is equivalent to the maxent Lagrangian, and in particular
has the same solution, Eq.~\ref{eq:boltzpr}.

Another alternative interpretation involves {\bf{world cost}}
functions, which are quantifications of the quality of a joint pure
strategy $x$ from the point of view of an external observer (e.g., a
system designer, the government, an auctioneer, etc.). A particular
class of world cost functions are ``social welfare functions'', which can
be expressed in terms of the cost functions of the individual
players. Perhaps the simplest example is $G(x) = \sum_i \beta_i
g_i(x)$, where the $\beta_i$ serve to trade off how much we value one
player's cost vs. anothers. If we know the value of this social
welfare function, but nothing else, then maxent tells us to minimize
the Lagrangian of Eq.~\ref{eq:firstlag}.

\subsection{Bounded rational game theory}
\label{sec:bounded}

In many situations we have prior knowledge different from (or in
addition to) expected values of cost functions. This is particularly
true when the players are human beings (so that behavioral economics
studies can be brought to bear) or simple computational algorithms. To
apply information theory in such situations, we simply need to
incorporate that prior knowledge into our Lagrangian(s).

To give a simple example, say that we know that the players all want
to ensure not just a low expected cost, but also that the actual cost
doesn't vary too much from one sample of $q$ to the next. We can
formalize this by saying that in addition to expected costs, our prior
knowledge includes variances in the costs. Given the expected values
of the costs, such variances are specified by the expected values of
the squares of the cost. Accordingly, all our prior knowledge is in
the form of expectation values. Modifying Eq.~\ref{eq:boltzpr}
appropriately, we arrive at the solution
\begin{equation}
q_i(x_i)  \propto e^{-E_{q_{(i)}}(\alpha_i (g_i - \lambda_i)^2 \mid x_i)}. \nonumber
\label{eq:gauss}
\end{equation}
\noindent where the Lagrange parameters $\alpha_i$ and $\lambda_i$ are
given by the provided expectations and variances of the costs of the
players.

Eq.~\ref{eq:gauss} is our best guess for what the actual mixed strategy
of player $i$ is, in light of our prior knowledge concerning that
player. Note that this formula directly reflects the fact that player
$i$ does not care only about minimizing cost, i.e., maximizing
utility.  In this, we are directly incorporating the possibility that
the player violates the axioms of utility theory --- something never
allowed in conventional game theory. Other behavioral economics
phenomena like risk aversion can be treated in a similar fashion.

A variant of this scenario would have our prior knowledge only give
the variances of the costs of the players and not their expected
costs. In this cost the Lagrangian must involve a term quadratic in
$q$, in addition to the entropy term and a term linear in $q$. (See
the subsection on multiple cost functions.) More generally, our prior
knowledge can be any nonlinear function of $q$. In addition, even if
we stick to prior knowledge that is linear in $q$, that knowledge can
couple the cost functions of the players. For example, if we know that
the expected difference in cost of players $i$ and $j$ is $\epsilon$,
the associated Lagrange constraint term is $\int dx q(x) [g_i(x) -
g_j(x) - \epsilon]$. In this situation our prior knowledge couples the
strategies of the players, even though those players are
independent. See the discussion on constrained optimization in
Sec. Opt.

\subsection{Cost of computation}

As mentioned above, bounded rationality is an unavoidable consequence
of the cost of computation to player $i$ of finding its optimal
strategy. Unfortunately, one cannot simply incorporate that cost into
$g_i$, and then presume that the player acts perfectly rationally for
this new $g_i$. The reason is that this cost is associated with the
entire distribution $q_i(x_i)$ that player $i$ calculates; it not
associated with some particular joint-strategy formed by sampling such
a distribution.

How might we quantify the cost of calculating $q_i$? The natural
approach is to use information theory. Indeed, that cost arises
naturally in the bounded rationality formulation of game theory
presented above. To see how, for each player $i$ define 
\begin{equation}
f_i(x, q_i(x_i)) \equiv \beta_i g_i(x) + {\mbox{ln}}
[q_i(x_i)]. \nonumber
\end{equation}
\noindent Then we can write the maxent Lagrangian for player $i$ as
\begin{equation}
L_i(q) = \int dx \; q(x) f_i(x, q_i(x_i)) . 
\label{eq:compcost}
\end{equation}
Now in a bounded rational game every player sets its strategy to
minimize its Lagrangian, given the strategies of the other players. In
light of Eq.~\ref{eq:compcost}, this means that we {\it{can}}
interpret each player in a bounded rational game as being perfectly
rational for a cost function that incorporates its computational
cost. To do so we simply need to expand the domain of ``cost
functions" to include probability values as well as joint moves.

Similar results hold for non-maxent Lagrangians. All that's needed is
that we can write such a Lagrangian in the form of
Eq.~\ref{eq:compcost} for some appropriate function $f_i$.

\subsection{Multiple cost functions per player}
\label{multiplecost}

Say player $i$ has several different cost functions \{$g^j_i$\} and
wants to choose a strategy that will do well at all of them. In the
case of pure strategies we can simply define an aggregate function
like ${\mbox{max}}_j g_i^j(x)$ or $\sum_j [g^j_i(x)]^2$, and employ
that in a conventional, single-cost-function-per-player game theoretic
analysis. Player $i$ will perform well according to such a function
iff it performs well according to all of the constituent $g^j_i$.

One might think that for mixed strategies one could just ``roll up"
the cost functions and say that player $i$ works to minimize an
aggregate cost function $\frac{\sum_j g^j_i}{\sum_j 1}$. However
especially when player $i$ has many cost functions, it may be that
performance according to one or more of the constituent cost functions
is quite bad even though the performance according to this average
function is good. Similarly, player $i$ can have a low value of the
expectation of the minimum of its cost functions, even though the
minimum of the expected costs is quite high. More generally, we cannot
ensure that $E_q(g^j_i) = \int dx \; g^j_i(x) q_i(x) q_{(i)}(x_{(i)})$
has a good value for all $j$ by appropriately defining an aggregate
$g_i$. Instead, we must ``redefine" expected cost.

We can address this by modifying our goals, in analogy with the goals
typically ascribed to players playing pure strategies. We do this by
having the choice of cost function for player $i$ be controlled by a
fictional player. For example, conventional game theory analyzes the
case where player $i$ chooses a pure strategy to minimize the worst
case (over other players' moves) cost to $i$, i.e., to minimize
max$_{x_{(i)}} g^i(x_i, x_{(i)}$. Here the analogy would be for the
player to choose a mixed strategy to minimize the worst case (over
moves by the fictional player) expected cost, i.e., to minimize
${\mbox{max}}_j E_q(g_i^j)$. A similar choice, appropriate when the
cost functions are all positive-definite, is for player $i$ to
minimize $\sum_j [E_q(g^j_i)]^2$.\footnote{Note that due to the
convexity of the squaring operator such minimization will help ensure
that no single expectation value $E_q(g^j_i)$ is too high. Minimizing
the expectation value $\int dx q(x) \sum_j [g^j_i(x)]^2$ will do a
roughly similar thing, in that it will help ensure that $q(x)$ is
small where the individual $g_j(x)$ are large.}  Formally, such
functions are just Lagrangians of $q$. If we wish, we can modify them
to incorporate bounded rationality, getting Lagrangians like $\sum_j
\beta_j [E_q(g^j_i)]^2 - S(q_i)$, where the $\beta_j$ determine the
relative rationalities of player $i$ according to its various cost
functions.

These kinds of Lagrangians can also model the process of mechanism
design, where there is an external designer who induces the players to
adopt a desirable joint-strategy~\cite{futi91}. As an example,
``desirable" sometimes means that no single player's expected cost is
high. A system that meets this goal fairly well can be modeled with
a Lagrangian involving terms like $\sum_i [E_q(g_i)]^2$.

\subsection{Shape of the Lagrangian surface}

To analyze the shape of the Lagrangian, we start with the following
lemma, which extends the technique of Lagrange parameters to
off-equilibrium points:

$ $

\noindent {\bf{Lemma 1:}} Consider the set of all vectors leading from $x' \in
{\mathbb{R}}^n$ that are, to first order, consistent with a set of
constraints over ${\mathbb{R}}^n$.  Of those vectors, the one giving
the steepest ascent of a function $V(x)$ is $\vec{u} = \nabla V +
\sum_i \lambda_i \nabla f_i$, up to an overall proportionality
constant, where the $\lambda_i$ enforce the first order consistency
conditions, $\vec{u} \cdot \nabla f_i = 0 \;\; \forall i$.

$ $

This lemma can be used to establish that at the edge of $\cal{Q}$,
the space of all product distributions $q$, the steepest descent
direction of any player's Lagrangian points into the interior of
$\cal{Q}$ (assuming finite $\beta$ and $\{g_i\}$). Accordingly, whereas
Nash equilibria can be on the edge of $\cal{Q}$ (e.g., for a pure
strategy Nash equilibrium), in bounded rational games any equilibrium
must lie in the interior of $\cal{Q}$. In other words, any equilibrium
(i.e., any local minimum) of a bounded rational game has non-zero
probability for all joint moves. So we never have to consider extremal
mixed strategies in searching for equilibria.

Lemma 1 can also be used to construct examples of games with more than
one bounded rational equilibrium (just like there are games with more
than Nash equilibrium). One can also show that for every player $i$
and any point $q$ interior to $\cal{Q}$, there are directions in
$\cal{Q}$ along which $i$'s Lagrangian is locally convex.
Accordingly, no player's Lagrangian has a local maximum interior to
$\cal{Q}$. So if there are multiple local minima of $i$'s Lagrangian,
they are separated by saddle points across ridges. Similarly, the
uniform $q$ is a solution to the set of coupled equations
Eq. \ref{eq:boltzpr} for a team game, but typically is not a local
minimum, and therefore must be a saddle point.

Say we modify the Lagrangians to be defined for all possible $p$, not
just those that are product distributions. For example the Lagrangian
of Eq.~\ref{eq:firstlag} becomes
\begin{equation}
L(p) \equiv \sum_i \beta_i [\int dx \; g_i(x) p(x) - \epsilon_i] -
S(p) . \nonumber
\end{equation}
The first term in this Lagrangian is linear in $p$. Since entropy is a
concave function of the Euclidean vector $p$ over the unit simplex,
this means that the overall Lagrangian is a convex function of $p$
over the space of allowed $p$. This means there is a {\it{unique}}
minimum of the Lagrangian over the space of all possible legal $p$.
Furthermore, as mentioned previously, for finite $\beta$ at least one
of the derivatives of the Lagrangian is negative infinite at the border
of the allowed region of $p$. This means that the unique minimum of
the Lagrangian is interior to that region, i.e., is a legal probability
distribution.

In general this optimal $p$ will not be a product distribution, of
course. Rather the strategy choices of the players are typically
statistically coupled, under this $p$. Such coupling is very
suggestive of various stochastic formulations of noncooperative game
theory. Coupling also arises in cooperative game theory, in which
binding contracts couple the moves of the players~\cite{grei99,auha92}.

Similarly, as in proven in the appendix, the Lagrangian $L(p) = \beta
\sum_i [E_p(g_i)]^2 - S(p)$ is convex over the manifold of legal $p$,
assuming non-negative $\beta$. So the model of mechanism design
introduced in Sec.~\ref{multiplecost} has a unique equilibrium --- if
we allow the players to be statistically coupled.

\subsection{Rationality operators}

Often our prior knowledge will not concern expected costs. In
particular, this is usually true if our prior knowledge is provided to
us before the game is played, rather than afterward. In such a
situation, prior knowledge will more likely concern the
``intelligences" of the players, i.e., how close they are to being
rational. In particular, if we want our prior knowledge concerning
player $i$ to be relatively independent of what the other players do,
we cannot use $i$'s expected cost as our prior knowledge. Our prior
knowledge will often concern how peaked $i$'s mixed strategy is about
whichever of its moves minimize its cost (or how peaked we can assume
it to be), not the associated minimal cost values.

Formally, the problem faced by player $i$ is how to set its mixed
strategy $q_i(x_i)$ so as to maximize the expected value of its {\bf
{effective cost function}}, $E(g_i \mid x_i)$.  Generalizing, what we
want is a {\it rationality} operator $R(U, p)$ that measures how
peaked an arbitrary distribution $p(y)$ is about the minimizers of an
arbitrary cost function $U(y)$, argmin${y} U(y)$.

Formally, we make two requirements of $R$:
\begin{enumerate}

\item If $p(y) \propto e^{-\beta U(y)}$, for non-negative $\beta$,
then it is natural to require that the peakedness of the distribution
--- its rationality value --- is $\beta$.

\item We also need to also specify something of $R(U, p)$'s
behavior for non-Boltzmann $p$. It will suffice to require that of the
$p$ satisfying $R(U, p) = \beta$, the one that has maximal entropy is
proportional to $e^{-\beta U(y)}$. In other words, we require that
the Boltzmann distribution maximizes entropy subject to a provided
value of the rationality operator.
\end{enumerate}
As an illustration, a natural choice for $R(U, p)$ would be the
$\beta$ of the Boltzmann distribution that ``best fits" $p$.
Information theory provides us such a measure for how well a
distribution $p_1$ is fit by a distribution $p_2$. This is the
{\bf{Kullback-Leibler distance}}~\cite{coth91,duha00}:
\begin{equation}
KL(p_1 \; || \; p_2) \equiv S(p_1 \; || \; p_2) - S(p_1)
\end{equation}
\noindent where $S(p_1 \; || \; p_2) \equiv -\int dy \; p_1(y)
{\mbox{ln}}[\frac{p_2(y)}{\mu(y)}]$ is known as the {\bf{cross
entropy}} from $p_1$ to $p_2$ (and as usual we implicitly choose
uniform $\mu$). The KL distance is always non-negative, and equals
zero iff its two arguments are identical.

Define $N(U) \equiv \int dy \; e^{-U(y)}$, the normalization constant for
the distribution proportional to $e^{-U(y)}$. (This is called the
{\bf{partition function}} in statistical physics.) Then using the KL
distance, we arrive at the rationality operator
\begin{eqnarray*}
R_{KL}(U, p) &\equiv& {\mbox{argmin}}_\beta KL(p \; || \; \frac{e^{-\beta
U}}{N(\beta U)}) \nonumber \\
&=& {\mbox{argmin}}_\beta [\beta \int dy\; p(y) U(y) + {\mbox
{ln}}(N(\beta U))] .
\end{eqnarray*}
\noindent In the appendix it is proven that $R_{KL}$ respects the two
requirements of rationality operators.

The quantity ${\mbox {ln}}(N(\beta U))$ appearing in the second
equation, when scaled by $\beta^{-1}$, is called the {\bf{free
energy}}. It is easy to verify that it equals the Lagrangian $E_p(U) -
S(p)/\beta$ if $p$ is given by the Boltzmann distribution $p(y)
\propto e^{-\beta U(y)}$.

Say our prior knowledge is \{$\rho_i$\}, the rationalities of the
players for their associated effective cost functions.  Introduce the
general notation 
\begin{eqnarray*}
[U]_{i,p}(x_i) \equiv \int dx_{(i)} U(x_i, x_{(i)})
p(x_{(i)} \mid x_i),
\end{eqnarray*}
\noindent so that $[g_i]_{i,q}$ is player $i$'s effective cost
function. Then the Lagrangian for our prior knowledge is
\begin{eqnarray} L(q) &=& \sum_i \lambda_i [R({[g_i]}_{i,q}, q_i) -
\rho_i] \; - \; S(q) .  \label{eq:firstratlag} \end{eqnarray}
\noindent where the $\lambda_i$ are the Lagrange parameters. Just as
before, there is an alternative way to motivate this Lagangian: if our
prior knowledge consists of the entropy of the joint system, and we
assume each player will have maximal rationality subject to that prior
knowledge, we are led to the Lagrangian of Eq.~\ref{eq:firstratlag}.

It is shown in the appendix that for the Kullback-Leibler rationality
operator, we can replace any constraint of the form $R({[g_i]}_{i,q},
q_i) = \rho_i$ with $E_q(g_i) = \int dx \; g_i(x) \frac{e^{-\rho_i
E(g_i \mid x_i)}}{N(\rho_i g_i)} q_{(i)}(x_{(i)})$. In other words,
knowing that player $i$ has KL rationality $\rho_i$ is equivalent to
knowing that the actual expected value of $g_i$ equals the ``ideal
expected value'', where $q_i$ is replaced by the Boltzmann
distribution of Eq.~\ref{eq:boltzpr} with $\beta = \rho_i$. This
contrasts with the prior knowledge underlying the Lagrangian in
Eq.~\ref{eq:firstlag}, in which we know the actual numerical value of
$E_q(g_i)$.

Just as before, we can focus on player $i$ by augmenting our prior
knowledge to include the strategies of all the other players. The
associated Lagrangian is
\begin{equation}
L_i(q_i) = \lambda_i [R([g_i]_{i,q}, q_i) - \rho_i] \; - \; S(q_i) .
\label{eq:secondratlag}
\end{equation}
\noindent (The prior knowledge concerning the strategies of the other
players is manifested in the effective cost function.)  It is shown in
the appendix that the  set of all the Lagrangians in
Eq.~\ref{eq:secondratlag} (one for each player) are minimized
simultaneously by any distribution of the form
\begin{equation}
q^g \equiv \frac{\prod_i e^{-\rho_i [g_i]_{i,q}}}{N(\rho_i [g_i]_{i,q})} \nonumber
\end{equation}
\noindent In addition, since this distribution obeys all the
constraints in the Lagrangian in Eq.~\ref{eq:firstratlag}, we know
that there exists a minimizer of that Lagrangian. All of this holds
regardless of the precise rationality operator one uses.

Note that the Lagrangian $L_i$ of Eq.~\ref{eq:secondratlag} for player
$i$ arises in response to prior knowledge specific to player
$i$. Changing from one player and its Lagrangian to another changes
the prior knowledge.  (The same is true for the Lagrangians in
Eq.~\ref{eq:maxentlag}.)  In contrast, the Lagrangian of
Eq.~\ref{eq:firstratlag} arises for a single unified body of prior
knowledge, namely the set of all players' rationalities.

For that single body of knowledge, the equilibrium of the game is the
solution to a {\it single}-objective optimization problem.  This
contrasts with the conventional formulation of full rationality game
theory, where the equilibrium is cast as a solution to a
multi-objective optimization problem (one objective per
player). Furthermore, for finite $\beta$, at least one of the
derivatives of the Lagrangian is negative infinite at the border of the
allowed region of product distributions (i.e., at the border of the
Cartesian product of unit simplices). Accordingly, all solutions lie
in the interior of that region. This can be a big advantage for
finding such solutions numerically, as elaborated below.

\subsection{Semi-coordinate systems}
\label{semicoord}

Consider a multi-stage game like chess, with the stages (i.e., the
instants at which one of the players makes a move) delineated by
$t$. Now strategies are what are set by the players before play
starts. So in such a multi-stage game the strategy of player $i$,
$x_i$, must be the set of $t$-indexed maps taking what that player has
observed in the stages $t' < t$ into its move at stage $t$. Formally,
this set of maps is called player $i$'s {\bf{normal form}}
strategy. 

The joint strategy of the two players in chess sets their joint
move-sequence, though in general the reverse need not be true. In
addition, one can always find a joint strategy to result in any
particular joint move-sequence. More generally, any onto mapping
$\zeta: x \rightarrow z$, not necessarily invertible, is called a
{\bf{semi-coordinate system}}.  The identity mapping $z \rightarrow z$
is a trivial example of a semi-coordinate system.  Another example is
the mapping from joint-strategies in a multi-stage game to joint
move-sequences is an example of a semi-coordinate system. So changing
the representation space of a multi-stage game from move-sequences $z$
to strategies $x$ is a semi-coordinate transformation of that game.

Typically there is overlap in what the players in chess have observed
at stages preceding the current one. This means that even if the
players' strategies are statistically independent, their move
sequences are statistically coupled.  In such a situation, by
parameterizing the space of joint-move-sequences $z$ with
joint-strategies $x$, we shift our focus from the coupled distribution
$P(z)$ to the decoupled product distribution, $q(x)$. This is the
advantage of casting multi-stage games in terms of normal form
strategies.

We can perform a semi-coordinate transformation even in a single-stage
game. Say we restrict attention to distributions over spaces of
possible $x$ that are product distributions.  Then changing $\zeta(.)$
from the identity map to some other function means that the players
are no longer independent. After the transformation their strategy
choices --- the components of $z$ --- are statistically coupled, even
though we are considering a product distribution.  

Formally, this is expressed via the standard rule for transforming
probabilities, 
\begin{eqnarray}
P_z(z) \equiv \zeta(P_x) \equiv \int dx P_x(x) \delta(z - \zeta(x)),
\label{eq:semitransf}
\end{eqnarray}
\noindent where $\zeta(.)$ is the mapping from $x$ to $z$, and $P_x$
and $P_z$ are the distributions across $x$-space and $z$-space,
respectively. To see what this rule means geometrically, let $\cal{P}$
be the space of all distributions (product or otherwise) over
$z$'s. Recall that $\cal{Q}$ is the space of all product distributions
over $x$, and let $\zeta({\cal{Q}})$ be the image of $\cal{Q}$ in
$\cal{P}$. Then by changing $\zeta(.)$, we change that image;
different choices of $\zeta(.)$ will result in different manifolds
$\zeta({\cal{Q}})$.

As an example, say we have two players, with two possible strategies
each. So $z$ consists of the possible joint strategies, labeled $(1,
1), (1, 2), (2, 1)$ and $(2, 2)$. Have the space of possible $x$ equal
the space of possible $z$, and choose $\zeta(1, 1) = (1, 1), \;
\zeta(1, 2) = (2, 2), \; \zeta(2, 1) = (2, 1)$, and $\zeta(2, 2) = (1,
2)$. Say that $q$ is given by $q_1(x_1 = 1) = q_2(x_2 = 1) =
2/3$. Then the distribution over joint-strategies $z$ is $P_z(1, 1) =
P_x(1, 1) = 4/9$, $P_z(2, 1) = P_z(2, 2) = 2/9$, $P_z(1, 2) = 1/9$. So
$P_z(z) \ne P_z(z_1)P_z(z_2)$; the strategies of the players are
statistically coupled.

Such coupling of the players' strategies can be viewed as a
manifestation of sets of potential binding contracts. To illustrate
this return to our two player example. Each possible value of a
component $x_i$ determines a pair of possible joint strategies. For
example, setting $x_1 = 1$ means the possible joint strategies are
$(1, 1)$ and $(2, 2)$. Accordingly such a value of $x_i$ can be viewed
as a set of proffered binding contracts. The value of the other
components of $x$ determines which contract is accepted; it is the
intersection of the proffered contracts offered by all the components
of $x$ that determines what single contract is selected. Continuing
with our example, given that $x_1 = 1$, whether the joint-strategy is
$(1, 1)$ or $(2, 2)$ (the two options offered by $x_1$) is determined
by the value of $x_2$.

Binding contracts are a central component of cooperative game
theory. In this sense, semi-coordinate transformations can be viewed
as a way to convert noncooperative game theory into a form of
cooperative game theory.  

While the distribution over $x$ uniquely sets the distribution over
$z$, the reverse is not true.  However so long as our Lagrangian
directly concerns the distribution over $x$ rather than the
distribution over $z$, by minimizing that Lagrangian we set a
distribution over $z$. In this way we can minimize a Lagrangian
involving product distributions, even though the  associated
distribution in the ultimate space of interest is not a product
distribution.

The Lagrangian we choose over $x$ should depend on our prior
information, as usual. If we want that Lagrangian to include an
expected value over $z$'s (e.g., of a cost function), we can directly
incorporate that expectation value into the Lagrangian over $x$'s,
since expected values in $x$ and $z$ are identical: $\int dz P_z(z) A(z)
= \int dx P_x(x) A(\zeta(x))$ for any function $A(z)$. (Indeed, this is
the standard justification of the rule for transforming probabilities,
Eq.~\ref{eq:semitransf}.)

However other functionals of probability distributions can differ
between the two spaces. This is especially common when $\zeta(.)$ is
not invertible, so the space of possible $x$ is larger than the space
of possible $z$. For example, in general the entropy of a $q \in
\cal{Q}$ will differ from that of its image, $\zeta(q) \in
\zeta({\cal{Q}})$ in such a case. (The prior probability $\mu$ in the
definition of entropy only gives us invariance when the two spaces
have the same cardinality.) A correction factor is necessary to relate
the two entropies.

In such cases, we have to be careful about which space we use to
formulate our Lagrangian.  If we use the transformation $\zeta(.)$ as
a tool to allow us to analyze bargaining games with binding contracts,
then the direct space of interest is actually the $x$'s (that is the
place in which the players make their bargaining moves). In such cases
it makes sense to apply all the analysis of the preceding sections
exactly as it is written, concerning Lagrangians and distributions
over $x$ rather than $z$ (so long as we redefine cost functions to
implicitly pre-apply the mapping $\zeta(.)$ to their
arguments). However if we instead use $\zeta(.)$ simply as a way of
establishing statistical dependencies among the strategies of the
players, it may make sense to include the entropy correction factor in
our $x$-space Lagrangian.

An important special case is where the following three conditions are
met: Each point $z$ is the image under $\zeta(.)$ of the same number
of points in $x$-space, $n$; $\mu(x)$ is uniform (and therefore so is
$\mu(z)$); and the Lagrangian in $x$-space, $L_x$, is a sum of
expected costs and the entropy.  In this situation, consider a
$z$-space Lagrangian, $L_z$, whose functional dependence on $P_z$, the 
distribution over $z$'s, is
identical to the dependence of $L_x$ on $P_x$, except that the
entropy term is divided by $n$
\footnote{For example, if $L_x(P_x) = \beta E_{P_x}(G(\zeta(.))) -
S(P_x)$, then $L_z(P_z) =
\beta E_{P_z}(G(.)) - S(P_z)/n$, where $P_x$ and $P_z$ are related
as in Eq.~\ref{eq:semitransf}.}. Now the minimizer $P^*(x)$ of $L_x$
is a Boltzmann distribution in values of the cost
function(s). Accordingly, for any $z$, $P^*(x)$ is uniform across all
$n$ points $x \in \zeta^{-1}(z)$ (all such $x$ have the same cost
value(s)).  This in turn means that $S(\zeta(P_x)) = nS(P_z)$ So our
two Lagrangians give the same solution, i.e., the ``correction factor"
for the entropy term is just multiplication by $n$.

\subsection{Entropic prior game theory}

Finally, it is worth noting that in the real world the information we
are provided concerning the system often will not consist of {\it
exact} values of functionals of $q$, be those values expected costs,
rationalities, or what have you. Rather that knowledge will be in the
form of data, $D$, together with an associated likelihood function
over the space of $q$. For example, that knowledge might consist of a
bias toward particular rationality values, rather than precisely
specified values: \begin{eqnarray} P(D \mid q) \propto e^{-\alpha
\sum_i [R_{KL}([g_i]_{i,q}) - \rho_i]^2} . \nonumber \end{eqnarray}
\noindent where $\alpha$ sets the strength of the bias.

The extension of the maximum entropy principle to such situations
uses the {\bf{entropic prior}}, $P(q) \propto e^{-\gamma
S(q)}$. Bayes' theorem is then invoked to get the posterior
distribution~\cite{jabr03}:
\begin{eqnarray}
P(q \mid D) \propto e^{-\sum_i \alpha_i [R_{KL}([g_i]_{i,q}) - \rho_i]^2
- \gamma S(q)}
. \nonumber
\end{eqnarray}
The {\bf{Bayes optimal}} estimate for $q$, under a quadratic penalty
term, is then given by $E(q \mid D)$.  The maxent principle for
estimating $q$ is given by this estimate under the limit of all
$\alpha_i$ going to infinity. For finite $\alpha$ solving for $E(q
\mid D)$ can be quite complicated though. For
simplicity, such cases are not considered here.

\section{PD theory and statistical physics}

There are many connections between bounded rational game theory --- PD
theory --- and statistical physics. This should not be too surprising,
given that many of the important concepts in bounded rational game
theory, like the Boltzmann distribution, the partition function, and
free energy, were first explored in statistical physics. This section
discusses some of these connections.

\subsection{Background on statistical physics}

Statistical physics is the physics of systems about which we have
incomplete information. An example is knowing only the  expected
value of a system's energy (i.e., its temperature) rather than the
precise value of the energy. The statistical physics of such systems
is known as the {\bf{canonical ensemble}}. Another example is the
{\bf{grand canonical ensemble}} (GCE). There the number of particles
of various types in the system is also uncertain. As in the canonical
ensemble, in the GCE what knowledge we do have takes the form of
expectation values of the quantities about which we are uncertain,
i.e., the number of particles of the various types that the system
contains, and the energy the system.

Traditionally these kinds of ensembles were analyzed in terms of
``baths" of the uncertain variable that are connected to the
system. For example, in the canonical ensemble the system is connected
to a heat bath. In the GCE the system is also connected to a bath of
particles of the various types.

Such analysis showed that for the canonical ensemble the probability
of the system being in the particular state $x$ is given by the
Boltzmann distribution over the associated value of the system's
energy, $G(x)$,  with $\beta$ interpreted as the (inverse) temperature
of the system: $p(x) \propto e^{-\beta G(x)}$. This result is
independent of the details characteristics of the physical system; all
that is important is the {\bf{Hamiltonian}} $G(x)$, and temperature
$\beta$.

Note that once one knows $p(x)$ and $G(x)$, one knows the expected
energy of the system. It is $G(x)$ that is a fixed property of the
system, whereas $\beta$ can vary. Accordingly, specifying $\beta$ is
exactly equivalent to specifying the expected energy of the system.

In the case of the GCE, $x$ implicitly specifies the number of
particles of the various types, as well as their precise state. The
analysis for that case showed that $p(x) \propto e^{-\beta G(x) -
\sum_i \mu_i n_i}$. In this formula $\beta$ is again the inverse
temperature, $n_i$ is the number of particles of type $i$, and $\mu_i
> 0$ is the {\bf{chemical potential}} of each particle of type $i$.

Jaynes was the first to show that these results of conventional
statistical physics could be derived without recourse to artificial
notions like ``baths", simply by using the maxent principle.  In
particular, he used the exact reasoning in Sec.~\ref{sec:bounded} to
derive the fact that the canonical ensemble is governed by the
Boltzmann distribution.

\subsection{Mean field theory and PD theory}

In practice it can be quite difficult to evaluate this Boltzmann
distribution, due to difficulty in evaluating the partition function.
For example, in a {\bf{spin glass}}, $x$ is an $N$-dimensional vector
of bits, one per particle, and $G(x) = \sum_{i,j} H_{i,j} x_i x_j$. So
the partition function is given by $\int dx e^{-\sum_{i,j} H_{i,j} x_i
x_j}$, where $H$ is a symmetric real-valued matrix, and as before we
use $\int$ to indicate the integral according to the appropriate
measure (here a point-sum measure). In general, evaluating this sum
for large numbers of spins cannot be done in closed form.

Mean Field (MF) theory is a technique for getting around this problem
by approximating the partition function.  Intuitively, it works by
treating all the particles as independent.  It does this by replacing
some of the values of the state of a particle in the Hamiltonian by
its average state. For example, in the case of the spin glass, one
approximates $\sum_{i,j} H_{i,j} [x_i - E(x_i)] [x_j - E(x_j)]
\approxeq 0$, where the expectation values are evaluated according to
the associated exact Boltzmann distribution, i.e., one assumes that
fluctuations about the means are relatively negligible. This then
means that
\begin{eqnarray*}
G(x) \approxeq \sum_{i,j} H_{i,j} 2 x_i E(x_j) \;-\; \sum_{i,j} H_{i,j}
E(x_i)E(x_j), 
\end{eqnarray*}
\noindent The second sum in this approximation cancels out when we
evaluate the associated approximate Boltzmann distribution, leaving us
with the distribution
\begin{eqnarray*}
p^{\beta U}(x) &\approxeq& P^{\beta U}(x) \equiv \frac{e^{-\beta \sum_{i,j} H_{i,j} 2 x_i E(x_j)}}{\int dx \;
e^{-\beta \sum_{i,j} H_{i,j} 2 x_i E(x_j)}} \\
&=& 
\prod_i \frac{e^{-\alpha_i x_i}}{\int dx_i \;
e^{-\alpha_i x_i}},
\end{eqnarray*}
\noindent where 
\begin{eqnarray*}
\alpha_i \equiv 2 \beta \sum_{j} H_{i,j} E(x_j).
\end{eqnarray*}

This approximation $P^{\beta U}(x)$ is far easier to work with than
the exact Boltzmann distribution, $p^{\beta U}(x) = \frac{e^{-\beta
G(x)}}{N(\beta U)}$, since each term in the product is for a single
spin by itself. In particular, if we adopt this approximation we can
use numerical techniques to solve the associated set of simultaneous
equations
\begin{eqnarray*}
E(x_i) = \frac{\partial} {\partial_{\alpha_i}} [\int dx_i \; e^{-
\alpha_i x_i}] \;\;\; \forall i
\end{eqnarray*}

\noindent for the $E(x_i)$ (so that those $E(x_i)$ are no
longer exactly equal to the expected values of the \{$x_i$\} under the 
distribution $p^{\beta U}(x)$). Given those $E(x_i)$ values, we can then
evaluate the associated approximate Boltzmann distribution explicitly.

The mean field approximation to the Boltzmann distribution is a product
distribution, and in fact is identical to the product distribution
$q^g$ of bounded rational game theory, for the team game where $g_i(x) =
2 \beta G(x) \; \forall i$. Accordingly, the ``mean field theory" approximation
for an arbitrary Hamiltonian $U$ can be taken to be the associated
team game $q^g$, which is defined for any $U$.

This bridge between bounded rational game theory and statistical
physics means that many of the powerful tools that have been developed
in statistical physics can be applied to bounded rational game
theory. They also mean that PD theoretic techniques can be applied in
statistical physics. In particular, it is shown elsewhere
\cite{wolp04b,mabi04} that if one replaces the identical cost function
of each player in a team game with different cost functions, then the
bounded rational equilibrium of that game can be numerically found far
more quickly. In the context of statistical physics, this means that
numerically solving for a MF approximation may be expedited by
assigning a different Hamiltonian to each particle.

\subsection{Information-theoretic misfit measures}

The proper way to approximate a target distribution $p$ with a
distribution from a set $\cal{C}$ is to first specify a misfit measure
saying how well each member of $\cal{C}$ approximates $p$, and then
solve for the member with the smallest misfit. This is just as true
when $\cal{C}$ is the set of all product distributions as when it is
any other set.

How best to measure distances between probability distributions is a
topic of ongoing controversy and research \cite{woma04a}. The most
common way to do so is with the infinite limit log likelihood of data
being generated by one distribution but misattributed to have come
from the other.  This is know as the {\bf{Kullback-Leibler
distance}}~\cite{coth91,duha00,mack03}:
\begin{equation}
KL(p_1 \; || \; p_2) \equiv S(p_1 \; || \; p_2) - S(p_1)
\end{equation}
\noindent where $S(p_1 \; || \; p_2) \equiv -\int dx \; p_1(x)
{\mbox{ln}}[\frac{p_2(x)}{\mu(x)}]$ is known as the {\bf{cross
entropy}} from $p_1$ to $p_2$ (and as usual we implicitly choose
uniform $\mu$). The KL distance is always non-negative, and equals
zero iff its two arguments are identical. However it it is far from
being a metric.  In addition to violating the triangle inequality, it
is not symmetric under interchange of its arguments, and in numerical
applications has a tendency to blow up. (That happens whenever the
support of $p_1$ includes points outside the support of $p_2$.)

Nonetheless, this is by far the most popular measure. It is
illuminating to use it as our misfit measure. As shorthand, define the
``$pq$ distance'' as $KL(p \; || \; q)$, and the ``$qp$ distance'' as
$KL(q \; || \; p$, where $p$ is our target distribution and $q$ is a
product distribution. Then it is straightforward to show that the $qp$
distance from $q$ to target distribution $p^{\beta U}$ is just the
maxent Lagrangian, up to irrelevant overall constants. In other words,
the $q$ minimizing the maxent Lagrangian --- the distribution arising
in MF theory --- is the $q$ with the minimal $qp$ distance to the
associated Boltzmann distribution.

However the $qp$ distance is the (infinite limit of the negative log
of) the likelihood that distribution $p$ would attribute to data
generated by distribution $q$. It can be argued that a better measure
of how well $q$ approximates $p$ would be based on the likelihood that
$q$ attributes to data generated by $p$. This is the $pq$ distance.
Up to an overall additive constant (of the canonical distribution's
entropy), the $pq$ distance is
\begin{eqnarray*}
KL(p \; || \; q) = -\sum_i \int dx \; p(x) {\mbox{ln}}[q_i(x_i)].
\end{eqnarray*}
\noindent  This is equivalent to a team game where each
coordinate $i$ has the ``Lagrangian''
\begin{eqnarray*}
L^*_{i}(q)
&\equiv& -\int dx_i \; p_i(x_i) {\mbox{ln}}[q_i(_i)] ,
\end{eqnarray*}
\noindent where $p_i(x_i)$ is the marginal distribution $\int dx_{(i)}
p(x)$.

The minimizer of this is just $q_i = p_i \; \forall i$, i.e., each
$q_i$ is set to the associated marginal distribution of $p$. So in
particular, when our target distribution is the canonical ensemble
distribution $p^{\beta U}$, the optimal $q$ according to $pq$ distance
is the set of marginals of $p^{\beta U}$. Note that unlike the
solution for $qp$ distance, here the solution for each $q_i$ is
independent of the $q_{(i)}$. So we don't have a game theory scenario;
we do not need to pay attention to the $q_{(i)}$ when estimating each
separate $q_i$. Correspondingly, whereas there are many local minima
of the team game Lagrangian studied above, $q \in {\cal{Q}}
\rightarrow KL(q \; || \; p^{\beta U})$, there is only one, global
minimum of $q \in {\cal{Q}} \rightarrow KL(p^{\beta} \; || \; q)$.

Another difference between the two kinds of KL distance is how the
associated optimal product distributions are typically calculated
numerically.  The product distribution that optimizes the maxent
Lagrangian is usually found via derivative-based traversal of that
Lagrangian, or techniques like (mixed) Brouwer updating\cite{wolp04b,
mabi04, biwo04,lewo04,aiwo04}. In contrast, the integral giving each
marginal distribution of $p$ is usually found via adaptive importance
sampling of the associated integral, with the proposal distribution
for the integral to approximate $p_i$ set adaptively, as
$q_{(i)}$\cite{wolp04b}.

It is possible to motivate yet other choices for the $q$ that best
approximates $p^{\beta U}$. To derive one of them, start with Lemma 1,
with ${\mathbb{R}}^n$ set to the space of real-valued functions over
the set of $x$'s (so that $n$ is the number of possible $x$).  Have a
single constraint $f$ that restricts us to $\cal{P}$, the unit simplex
in ${\mathbb{R}}^n$, i.e., that restricts us to the set of functions
that (assuming they are nowhere-negative) are probability
distributions. Choose $V$ to be the associated Lagrangian, $L(p) =
\beta E_p(G) - S(p)$, $p$ being a point in our constrained submanifold
of ${\mathbb{R}}^n$. Note that this $p$ can be {\it{any}} distribution
over the $x$'s, including one that couples the components $\{x_i\}$.

Say we are at some current product distribution $q$. Then we can apply
Lemma 1 with the choices just outlined to tell us what direction to
move from $q$ in $\cal{P}$ so as to reduce the Lagrangian. In general,
taking a step in that direction will result in a distribution $p'$
that is not a product distribution. However we can solve for the
product distribution that is closest to that $p'$, and move to that
product distribution. By iterating this procedure we can define a
search over the submanifold of product distributions. We can then
solve for the product distribution at which this search will
terminate.

To do this, of course, we must define what we mean by ``closest''. Say
that we choose to measure closeness by $pq$ distance.  Then
the terminating production distribution is the one for which the
marginals of $\nabla L + \lambda \nabla f$ all equal 0. For each $i$,
this means that
\begin{eqnarray*}
&&\int dx_{(i)} [\beta G(x) + {\mbox{ln}}(p(x)) + 1 + \lambda] = 0
\end{eqnarray*}
at the equilibrium product distribution $p$. Writing out $p = \prod_i
q_i$ and evaluating gives
\begin{eqnarray}
&&q_i(x_i) \propto \exp{(-\beta{\frac{\int dx_{(i)} G(x)}{\int dx_{(i)}
	1}})}.
\end{eqnarray}

\noindent This is akin to the $q^g$ of a bounded rational game, except
that each player/particle $i$ sets its distribution by evaluating
conditional expected $U$ with a uniform distribution over the
$x_{(i)}$, rather than with $q_{(i)}$.

\subsection{Semi-coordinate transformations}

Let's say there are numerical difficulties with our finding a $q$ that
is local minimization of the maxent Lagrangian. That $q$ might still
be a poor fit to $p(x)$ if it is far from the global minimizer of the
Lagrangian. Furthermore, even the global minimizer might be a poor
fit, if $p(x)$ simply can't be well-approximated by a product
distribution.

There are many techniques for improving the fit of a product
distribution to a target distribution in machine learning and
statistics~\cite{duha00}. To give a simple example, say one wishes to
approximate the target distribution in ${\mathbb{R}}^N$ with a product
of Gaussians, one Gaussian for each coordinate. Even if the target
distribution a Gaussian, if it is askew, then one won't be able to do
a good job of approximating it with a product of Gaussians. However
one can use Principal Components Analysis (PCA) to find how to rotate
one's coordinates so that a product of Gaussians fits the target
exactly.

Similar techniques can address both the issue of breaking free of
local minima of the Lagrangian, and improving the accuracy of the best
product distribution approximation to $p$. More precisely, identify
$x$ with the variables $z$ discussed in Sec. \ref{semicoord}.  Then
consider changing the map $\zeta(.) : x \rightarrow z$ from the
identity map. This will in general change the mapping from $P_x$ to 
$L_z(\zeta(P_x))$. So if $L_z$ is the Lagrangian we are interested in, 
the mapping from product distributions over $x$ can be changed by
changing $\zeta(.)$, in general. 

As an example, consider the case where the space of $x$'s is identical
to the space of $z$'s, and consider all possible bijective
transformations $\zeta(.)$. Entropy is the same in both spaces for any
$\zeta$, i.e., $S(P_z) = S(\zeta(P_x)) = S(P_x)$. So for fixed $P_x$,
the entropy in $z$-space is independent of $\zeta(.)$. However if we
fix $P_x$ and change $\zeta(.)$ the expected values of utilities will
change. So $L_z(\zeta(P_x))$ does depend on $\zeta(.)$, as claimed.

This means that by changing $\zeta(.)$ while leaving $q_x$ unchanged,
we will in general change whether we are at a local minimum of
$L_z(\zeta(q_x))$. Furthermore, such a change will change how closely
the global minimizer of $L_z(\zeta(q_x))$ approximates any particular
target distribution.  Indeed, some such transformation will always
transform a team game to have a strictly convex maxent Lagrangian,
with only one (bounded rational) equilibrium, an equilibrium that is
in the interior of the region of allowed $q$ and that has the lowest
possible value of the Lagrangian. In the worst case, we can get this
behavior by transforming to the semi-coordinate system in which $x$ is
one-dimensional, so that any $p(z)$ --- coupling its variables or not
--- can be expressed as a $q(x) = q_1(x_1)$.

Note that unlike with PCA, semi-coordinate transformations can be used
for non-Euclidean semi-coordinates (i.e., when neither $x$'s nor $z$'s
are Euclidean vectors). They also can be guided by numerous measures
of the goodness of fit to the target distribution (e.g., KL distance),
in contrast to PCA's restriction to assuming a Gaussian likelihood.

\subsection{Bounded rational game theory for variable number of players}

The bridge between statistical physics and bounded rational game
theory have many uses beyond the practical ones alluded to the
previous subsection. In particular, it suggests extending bounded
rational game theory to ensembles other than the canonical ensemble.
As an example, in the GCE the number of particles of the various
allowed types is uncertain and can vary. The bounded rational game
theory version of that ensemble is a game in which the number of
players of various types can vary.

We can illustrate this by extending a simple instance of evolutionary
game theory~\cite{auha92} to incorporate bounded rationality and allow
for a finite total number of players. Say we have a finite population
of players, each of which has one of $m'$ possible
{\bf{types}}. (These are sometimes called {\bf{feature vectors}} in
the literature.) Each player $i$ in the population is randomly paired
with a different player $j$, and they each choose a strategy for a
two-person game. The set of strategies each of those players can
choose among is fixed by its respective attribute vector. In addition
the cost player $i$ receives depends on the attribute vectors of
itself and of $j$, in addition to their joint strategy. Finally, to
reflect this dependence, we allow each player to vary its strategy
depending on the attribute vector of its opponent; we call player
$i$'s {\bf{meta-strategy}} the mapping from its opponent's attribute
vector to $i$'s strategy.  \footnote{Note that it is trivial to
replace meta-strategies with strategies throughout the analysis below:
simply restrict attention to meta-strategies that do not vary with the
opponent's attribute vector}.

We encode an instance of this scenario in an $x$ with a countably
infinite number of dimensions. $x_{i,0} \equiv n_i(x)$ specifies the
number of players of type $i$, with $\vec{n}(x)$ being the vector of
the number of players of all types. For $1 < j \le x_{i,0}$, $x_{i,j}
\equiv s_{i,j}(x)$ the meta-strategy selected by the $j$'th player of
type $i$. If its opponent is the $j$'th player of type $T'$, the cost
to the $i$'th player of type $T$ is $g_{T,i,T',j}(x) \equiv
g_{T,i,T',j}(s,s',n_T,n_{T'})$, where $s$ and $s'$ are the two
players' respective meta-strategy. To enforce consistency between the
index numbers $i, j$ and the associated numbers of players, we set
$g_{T,i,T',j}(s,s',\vec{n})= 0$ if either $i > n_T$ or $j >
n_{T'}$.

To start we parallel the GCE, and presume that for each type we know
the expected number of players having that type, and the expected cost
averaged over all players having that type. Also stipulate that the
distribution over $x$ is a product distribution, $q$. Then our prior
information specifies the values of

$\sum_{k>0} k \; q_{_{T,0}}(k) = \sum_{x_{_{T,0}}} x_{_{T,0}} \; q_{_{T,0}}(x_{_{T,0}})$

$ $

\noindent and

$ $

$\sum_{\vec{n}:n_{_{T}} > 0} q(\vec{n}) \sum_{_{T':n_{_{T'}}>0}}
[\frac{n_{T'}} {\sum_{T''} n_{T''}}] \sum_{j,k} \int ds_{_T}
ds_{_{T'}}$
\begin{eqnarray*}
&& [1 - \delta_{_{T,T'}} \delta_{_{j,k}}]
\frac{
q_{_{T,j}}(s_{_T}) q_{_{T',k}}(s_{_{T'}})
g_{_{T,j,T',k}}(s_{_T},s_{_{T'}},\vec{n})} {n_{_T}
n_{_{T'}} }
\end{eqnarray*}

\noindent $=$

$ $

$\sum_{x_{_{1,0}}} \ldots \sum_{x_{_{T,0}>0}} \ldots \sum_{x_{_{m',0}}}
\sum_{_{T'}} \sum_{j,k} \int dx_{_{T,j}} dx_{_{T',k}}$
\begin{eqnarray*}
&& \{  [1 - \delta_{_{T,T'}}
\delta_{_{j,k}}] \; [\prod_{i=1}^{m'} q_{i,0}(x_{i,0})]  \;\;
\times \\
&& \;\;\;\;\;\;\;\;\; \frac{
q_{_{T,j}}(x_{_{T,j}}) \; q_{_{T',k}}(x_{_{T',k}})
\; g_{_{T,j,T',k}}(x)} {x_{_{T,0}} \;
{\sum_{T''}x_{_{T'',0}}} } \}
\end{eqnarray*}
\noindent respectively, for all types $T$. (The sums over $j$ and $k$
all implicitly extend from 1 to $\infty$, and the delta functions are
Kronecker deltas that prevent a player from playing itself.)

We can write these expressions as expectation values, over $x$, of $2m'$
functions. These functions are the $m'$ functions $n_T(x) = x_{T,0}$
(one function for each $T$) and the $m'$ functions
\begin{eqnarray*}
&&c_T(x) \equiv \frac{ \sum_{T',j,k}
\{[1 - \delta_{_{T,T'}}
\delta_{_{j,k}}] \; g_{_{T,j,T',k}}(x)\} }
{x_{_{T,0}} \; {\sum_{T''}x_{_{T'',0}}} } \Theta(x_{_{T,0}})
\end{eqnarray*}
\noindent respectively, where $\Theta$ is the Heaviside theta function 
that equals 1 if its argument exceeds 0, and equals 0 otherwise.
Accordingly, the maxent principle directs us to minimize the Lagrangian
\begin{equation*}
L(q) = -\sum_T[\mu_T (E(n_T) - N_T) + \beta_T(E(c_T) - C_T)] \;-\; S(q)
\end{equation*}
\noindent where the integers \{$N_T$\} and real numbers \{$C_T$\} are
our prior information. In the usual way, the solution for each pair $(i \in
\{1, \ldots, m'\}, j \ge 0)$ is
\begin{eqnarray*}
q_{_{i,j}}(x_{_{i,j}}) \propto e^{-E([\sum_{T'} \mu_{_{T'}} n_{_{T'}}
- \; \beta_{_{T'}}
c_{_{T'}}] \; \mid \; x_{_{i,j}})},
\label{eq:gce}
\end{eqnarray*}
\noindent where the values of the Lagrange parameters are all set by
our prior information.

This distribution is analogous to the one in the GCE.  As usual, one
can consider variants of it by focusing on one variable at a time,
having prior knowledge in the form of rationality values, etc.  In
addition, even if we stay in this random-2-player games scenario,
there is no reason for us to restrict attention to prior information
paralleling that of the GCE. As with bounded rational game theory with
a fixed number of players, our prior information can concern nonlinear
functions of $q$, couple the cost functions, etc.

In particular, in evolutionary game theory we do not know the expected
number of players having each type, nor their average costs. In
addition, the equilibrium concept stipulates that all players will
have type $T$ if a particular condition holds. That condition is that
the addition of a player of type other than $T$ to the population
results in an expected cost to that added player that is greater than
the associated expected cost to the players having type $T$. This
provides a model of the phenotypic interactions underlying natural
selection.

We can encapsulate evolutionary game theory in a Lagrangian by
appropriately replacing each pair of GCE-type constraints (one pair
for each type) with a single constraint. As an example, we
could have the (single) constraint for type $T$ be that
\begin{eqnarray}
E(\frac{n_{_T}}{\sum_{_{T'}} n_{_{T'}}}) =
E([\frac{{\mbox{max}}_{_{T'}}
c_{_{T'}} - c_{_T}}{{\mbox{max}}_{_{T'}}(c_{_{T'}}) -
{\mbox{min}}_{_{T'}}(c_{_{T'}})}]^{^\gamma})
\label{eq:evolgtconst}
\end{eqnarray}
\noindent for some positive real value $\gamma$.  For finite $\gamma$,
the entropy term in the Lagrangian ensures that for no $T$ is the
expectation value in the lefthand side of this constraint exactly
0.

In the limit of infinite $\gamma$, the distribution minimizing this
Lagrangian is non-infinitesimal only for the {\bf{evolutionarily
stable strategies}} of conventional evolutionary game theory. These
are the (type, strategy) pairs that are best performing, in the sense
that no other pair has a lower cost function value.  The distribution
for finite $\gamma$ can be viewed as a ``bounded rational'' extension
of conventional evolutionary game theory. In that extension (type,
strategy) pairs are allowed even if they don't have the lowest
possible cost, so long as their cost is close to the lowest possible
\footnote{Many other parameterized constraints will result in this
kind of relation between the parameter value and the resultant
Lagrangian-minimizing distribution. The one in
Eq.~\ref{eq:evolgtconst} was chosen simply for pedagogical clarity.}.

There is always a solution to this Lagrangian (unlike the case in
conventional full rationality evolutionary game theory). The technique
of Lagrange parameters provides that solution for each pair $(i \in
\{1, \ldots, m'\}, j \ge 0)$ in the usual way:
\begin{eqnarray*}
q_{_{i,j}}(x_{_{i,j}}) \propto e^{-E(\sum_{_{T'}} \alpha_{_{T'}} f_{_{T'}}(x) \; \mid \; x_{_{i,j}})}
\end{eqnarray*}
\noindent where the Lagrange parameters enforce our constraint, and
\begin{eqnarray*}
f_{_{T'}}(x) \equiv 
\frac{n_{_{T'}}}{\sum_{_{T''}} n_{_{T''}}}   - 
[\frac{{\mbox{max}}_{_{T''}} c_{_{T''}} - c_{_{T''}}} 
  {{\mbox{max}}_{_{T''}}(c_{_{T''}}) -
  {\mbox{min}}_{_{T''}}(c_{_{T''}})}]^{^\gamma} .
\end{eqnarray*}

More general forms of evolutionary game theory allow games with more
than two players, and localization via network structures delineating
how players are likely to be grouped to play a game. Other
elaborations have each player not know the exact attribute vectors of
all its opponents, but only an ``information structure" providing some
information about those opponents' attribute vectors. All such
extensions can be straightforwardly incorporated into the current
analysis. Many other extensions are simple to make as well. For
example, since the cost functions have all components of $\vec{n}$ in
their argument lists, they can depend on the total size of the
population. This allows us to model the effect on population size of
finite environmental resources.

Note that if we change how we encode the number of players of the
various types and their joint meta-strategy in $x$, we change the form of
the expectations in Eq.~\ref{eq:gce}. This reflects the fact that by
changing the encoding we change the implication of using a product
distribution. Formally, such a change in the encoding is a change in
the semi-coordinate system. See Sec.~\ref{semicoord}.

\section{Appendix}

This appendix provides proofs absent from the main text.

\subsection{$\beta \sum_i [E_p(g_i)]^2 - S(p)$ is convex over the unit simplex}

{\bf{Proof:}} Since $S(p)$ is concave over the unit simplex, and the
unit simplex is a hyperplane, it suffices to prove that $\sum_i
[E_p(g_i)]^2$ is convex over all of Euclidean space. Since a weighted
average of convex functions is convex, we only need to prove that any
single function of the form $[\int dx \; p(x) f(x)]^2$ is convex. The
Hessian of this function is $2 f(x) f(x')$. Rotate coordinates so that
$f$ is a basis vector, i.e., so that $f$ is proportional to a delta
function. This doesn't change the eigenvalues of the Hessian. After
this change though, the Hessian is diagonal, with one non-zero entry
on the diagonal, which is non-negative.  So its eigenvalues are zero
and a non-negative number. {\bf QED.}

\subsection{$R_{KL}$ is a rationality operator}

{\bf{Proof:}} Since KL distance only equals 0 when its arguments match
and is never negative, requirement (1) of rationality operators holds
for $R_{KL}$. Next, since $R_{KL} = {\mbox{argmin}}_\beta [\beta \int
dy\; p(y) U(y) + {\mbox {ln}}(N(\beta U))]$, we know that $E_p(U) =
-\frac{1}{N(\beta U)} \frac{\partial N(\beta U)}{\partial
\beta}|_{\beta = R_{KL}(U, p)}$. Accordingly, all $p$ with the same
rationality have the same expected value $E_p(U)$. Using the technique
of Lagrange parameters then readily establishes that of those
distributions having the same expected $U$, the one with maximal
entropy is a Boltzmann distribution. Furthermore, by requirement (1),
we know that for a Boltzmann distribution the exponent $\beta$ must
equal the rationality of that distribution.  {\bf QED.}

\subsection{Alternative form of a constraint on $R_{KL}$}

{\bf{Proof:}} Let $f\{\alpha, v\}$ be any function that is monotonically
decreasing in its (real-valued) first argument. Then any constraint
$R([g_i]_{i,q}, q_i) - \rho_i = 0$ is satisfied iff the constraint
$f\{R([g_i]_{i,q}, q_i), q_{(i)}\} - f\{\rho_i, q_{(i)}\} = 0$ is
satisfied. Choose
\begin{eqnarray*}
f\{\alpha, q_{(i)}\} &=& -\frac{\partial
{\mbox{ln}}(N(\beta[g_i]_{i,q}))}{\partial \beta}|_{\beta = \alpha} \\ \nonumber
&=& \frac{\int dx_i [g_i]_{i,q} e^{-\alpha[g_i]_{i,q}(x_i)}}{N(\alpha
[g_i]_{i,q})} . \nonumber
\end{eqnarray*}
\noindent Differentiating this quantity with respect to $\alpha$ gives
the negative of the variance of $[g_i]_{i,q}$ under the Boltzmann
distribution $\frac{e^{-\alpha [g_i]_{i,q}}}{N(\alpha
[g_i]_{i,q})}$. Since variances are non-negative, this derivative is
non-positive, which establishes that $f$ is monotonically decreasing
in its first argument.

Evaluating, 
\begin{eqnarray*}
f\{\rho_i, q_{(i)}\} = \int dx \; g_i(x) \frac{e^{-\rho_i
E(g_i \mid x_i)}}{N(\rho_i g_i)} q_{(i)}(x_{(i)}).
\end{eqnarray*}
\noindent In addition, from the equation defining $R_{KL}$, we know
that
\begin{eqnarray*}
-\frac{{\mbox{ln}}(N(\beta U(x_i)))}{\partial \beta}|_{\beta =
R_{KL}(U,q_i)} = \int dx_i q_i(x_i) U(x_i)
\end{eqnarray*}
\noindent for any function $U$. Plugging in $U = [g_i]_{i,q}$, we see
that
\begin{eqnarray*}
f\{R([g_i]_{i,q},q_i), q_{(i)}) &=& \int dx_i q_i(x_i) [g_i]_{i,q}(x_i) \\
&=& E_q(g_i). {\bf QED.}
\end{eqnarray*}

\subsection{$q^g$ minimizes the Lagrangians of Eq.~\ref{eq:secondratlag}}

{\bf{Proof:}} Following Nash, we can use Brouwer's fixed point theorem
to establish that for any non-negative \{$\rho_i$\}, there must exist
at least one product distribution given by $q^g$. The constraint term
in all the $L_i$ of Eq.~\ref{eq:secondratlag} is zero for this
distribution. By requirement (2), we also know that given $q^g_{(i)}$
(and therefore $[g_i]_{i,q^g}$), there is no $q_i$ with rationality
$\rho_i$ that has lower entropy than $q^g_i$.  Accordingly, no $q_i$
will have a lower value of $L_i$.  Since this holds for all $i$, $q^g$
minimizes all the Lagrangians in Eq.~\ref{eq:secondratlag}
simultaneously. {\bf{QED.}}

\subsection{Derivation of Lemma 1}

{\bf{Proof:}} Consider the set of $\vec{u}$ such that the directional
derivatives $D_{\vec{u}}f_i$ evaluated at $x'$ all equal 0. These are
the directions consistent with our constraints to first order. We need
to find the one of those $\vec{u}$ such that $D_{\vec{u}}g$ evaluated
at $x'$ is maximal.

To simplify the analysis we introduce the constraint that $|\vec{u}| =
1$. This means that the directional derivative $D_{\vec{u}}V$ for any
function $V$ is just $\vec{u} \cdot \nabla V$. We then use Lagrange
parameters to solve our problem. Our constraints on $\vec{u}$ are
$\sum_j u_j^2 = 1$ and $D_{\vec{u}}f_i(x') = \vec{u} \cdot \nabla
f_i(x') = 0 \; \; \forall i$.  Our objective function is
$D_{\vec{u}}V(x') = \vec{u} \cdot \nabla V(x')$.

Differentiating the Lagrangian gives
\begin{eqnarray*}
2 \lambda_0 u_i + \sum_i \lambda_i \nabla f = \nabla V \;\; \forall i.
\end{eqnarray*}
\noindent with solution 
\begin{eqnarray*}
u_i = \frac{\nabla V - \sum_i \lambda_i \nabla f}{2 \lambda_0}.
\end{eqnarray*}
\noindent $\lambda_0$ enforces our constraint on $|\vec{u}|$. Since we
are only interested in specifying $\vec{u}$ up to a proportionality
constant, we can set $2 \lambda_0 = 1$. Redefining the Lagrange
parameters by multiplying them by $-1$ then gives the result
claimed. {\bf QED.}

\subsection{Proof of claims following Lemma 1}

\noindent {\bf{i)}} Define $f_i(q) \equiv \int dx_i q_i(x_i)$, i.e.,
$f_i$ is the constraint forcing $q_i$ to be normalized.  Now for any
$q$ that equals zero for some joint move there must be an $i$ and an
$x'_i$ such that $q_i(x'_i) = 0$.  Plugging into Lemma 1, we can
evaluate the component of the direction of steepest descent along the
direction of player $i$'s probability of making move $x'_i$:
\begin{eqnarray*}
&&\frac{\partial L_i}{\partial q_i(x_i)} + \lambda \frac{\partial
  f_i}{\partial q_i(x_i)} = \\
&&\;\;\;\; \beta E(g_i \mid x_i) + {\mbox{ln}}(q_i(x_i)) - \frac{\int dx''_i [\beta
  E(g_i \mid x''_i) + {\mbox{ln}}(q_i(x''_i))]}{\int dx''_i 1}
\end{eqnarray*}

\noindent Since there must some $x''_i$ such tha $q_i(x''_i) \ne 0$,
$\exists x_i$ such that $\beta E(g_i \mid x''_i) +
{\mbox{ln}}(q_i(x''_i))$ is finite.  Therefore our component is
negative infinite. So $L_i$ can be reduced by increasing
$q_i(x'_i)$. Accordingly, no $q$ having zero probability for some
joint move $x$ can be a minimum of $i$'s Lagrangian.

$ $

\noindent {\bf{ii)}} To construct a bounded rational game with
multiple equilibria, note that at any (necessarily interior) local
minimum $q$, for each $i$, 
\begin{eqnarray*}
&& \beta E(g_i \mid x_i) + {\mbox{ln}}(q_i(x_i)) = \\
&& \;\;\;\;\;\;\;\; \beta \int dx_{(i)} g_i(x_i,
x_{(i)}) \prod_{j\ne i} q_j(x_j) + {\mbox{ln}}(q_i(x_i))
\end{eqnarray*}
\noindent must be independent of $x_i$, by Lemma 1. So say there is a
component-by-component bijection $T(x) \equiv (T_1(x_1), T_2(x_2),
\ldots)$ that leaves all the $\{g_j\}$ unchanged, i.e., such that
$g_j(x) = g_j(T(x)) \; \forall x, j$ \footnote{As an example, consider
a congestion team game in which all players have the same set of
possible moves, $G$ being a function only of the bit string indexed by
$k \in \mathbb{N}$, $\{N(x, k)\}$, where $N(x, k) = 1$ iff there is
a move that is shared by exactly $k$ of the players when the joint
move is $x$. In this case $T$ just permutes the set of possible moves
in the same way for all players.}.

Define $q'$ by $q'(x) = q(T(x)) \; \forall x$. Then for any two values 
$x^1_i$ and $x^2_i$,
\begin{eqnarray*}
&&\beta E_{q'}(g_i \mid x^1_i) + {\mbox{ln}}(q'_i(x^1_i))  \\
&& \;\;\;\;\;\;\;\;\;\;\;\;\;- \;\beta E_{q'}(g_i \mid x^2_i) \;+\; {\mbox{ln}}(q'_i(x^2_i)) \\
&&\;\;\;\;\;\;\;\;\;\;\;\;\;\;\;\;\;\;\;\;\;\;= \\
&&\beta \int dx_{(i)}  g_i(x^1_i, x_{(i)}) \prod_{j\ne i} q_j(T(x_j)) \;+\;
		{\mbox{ln}}(q_i(T(x^1_i))) \\
&&\;\;\;\;\; - \; \beta \int dx_{(i)}  g_i(x^2_i, x_{(i)})
			\prod_{j\ne i} q_j(T(x_j))) \;+\;
		{\mbox{ln}}(q_i(T(x^2_i))) \\
&& \;\;\;\;\;\;\;\;\;\;\;\;\;\;\;\;\;\;\;\;\;\;= \\
&&\beta \int dx_{(i)}  g_i(x^1_i, T^{-1}(x_{(i)})) \prod_{j\ne i} q_j(x_j) \;+\;
		{\mbox{ln}}(q_i(T(x^1_i))) \\
&&\;\;\;\;\; - \; \beta \int dx_{(i)}  g_i(x^2_i, T^{-1}(x_{(i)}))
			\prod_{j\ne i} q_j(x_j)) \;+\;
		{\mbox{ln}}(q_i(T(x^2_i))) \\
&& \;\;\;\;\;\;\;\;\;\;\;\;\;\;\;\;\;\;\;\;\;\;= \\
&&\beta \int dx_{(i)}  g_i(T(x^1_i), x_{(i)})) \prod_{j\ne i} q_j(x_j) \;+\;
		{\mbox{ln}}(q_i(T(x^1_i))) \\
&&\;\;\;\;\; - \; \beta \int dx_{(i)}  g_i(T(x^2_i), x_{(i)}))
			\prod_{j\ne i} q_j(x_j)) \;+\;
		{\mbox{ln}}(q_i(T(x^2_i))) \\
&& \;\;\;\;\;\;\;\;\;\;\;\;\;\;\;\;\;\;\;\;\;\;= \\
&&\beta E_{q}(g_i \mid T(x^1_i)) \;+\; {\mbox{ln}}(q_i(T(x^1_i)))  \\
&& \;\;\;\;\;\;\;\;\;\;\;\;\;- \;\beta E_{q}(g_i \mid T(x^2_i))
	\;+\; {\mbox{ln}}(q_i(T(x^2_i))) 
\end{eqnarray*}
where the invariance of $g_i$ was used in the penultimate step. Since
$q$ is a local minimum though, this last difference must equal
0. Therefore $q'$ is also a local minimum.

Now choose the game so that $\forall i, x_i, T(x_i) \ne x_i$. (Our
congestion game example has this property.) Then the only way the
transformation $q \rightarrow q(T)$ can avoiding producing a new
product distribution is if $q_i(x_i) = q_i(x'_i) \; \forall i, x_i,
x'_i$, i.e., $q$ is uniform. Say the Hessians of the players'
Lagrangians are not all positive definite at the uniform $q$. (For
example have our congestion game be biased away from uniform
multiplicities.) Then that $q$ is not a local minimum of the
Lagrangians. Therefore at a local minimum, $q \ne q(T)$. Accordingly,
$q$ and $q(T)$ are two distinct equilibria.

$ $

\noindent {\bf{iii)}}  To establish that at any $q$ there is always a
direction along which any player's Lagrangian is locally convex, fix
all but two of the $\{q_i\}$, $q_0$ and $q_1$, and fix both $q_0$ and
$q_1$ for all but two of their respective possible values, which we
can write as $q_0(0), q_0(1), q_1(0)$, and $q_1(1)$, respectively. So
we can parameterize the set of $q$ we're considering by two real
numbers, $x \equiv q_0(0)$ and $y \equiv q_1(0)$. The $2 \times 2$
Hessian of $L_i$ as a function of $x$ and $y$ has the entries
\begin{eqnarray*}
&& \;\; \frac{1}{x} + \frac{1}{a-x} \;\;\;\;\;\;\;\;\;\;\; \alpha \\
&& \;\;\;\;\;\;\;\;\; \alpha \;\;\;\;\;\;\;\;\;\;\;\; \frac{1}{y} + \frac{1}{b-y}
\end{eqnarray*}
\noindent where $a \equiv 1 - q_0(0) - q_0(1)$ and $b \equiv 1 -
q_1(0) - q_1(1)$, and $\alpha$ is a function of $g_i$ and $\prod_{j
\ne 0,1} q_j$. Defining $s \equiv \frac{1}{x} + \frac{1}{a-x}$ and 
$t \equiv \frac{1}{y} + \frac{1}{b-y}$, the eigenvalues of that
Hessian are 
\begin{eqnarray*}
\frac{s + t \pm \sqrt{4 \alpha^2 + (s - t)^2}}{2} .
\end{eqnarray*}
The eigenvalue for the positive root is necessarily
positive. Therefore along the corresponding eigenvector, $L_i$ is
convex at $q$. {\bf{QED.}}


\end{document}